\documentclass[aps,prb,superscriptaddress,floatfix,longbibliography,10pt]{revtex4-2}

\usepackage{amsmath,amssymb} 
\usepackage{bm} 
\usepackage{graphicx} 
\usepackage{comment} 
\usepackage{textcomp} 

\usepackage{enumitem}
\setlist{noitemsep}

\usepackage{xcolor} 
\definecolor{lightgray}{gray}{0.6}
\definecolor{medgray}{gray}{0.4}

\usepackage{hyperref}
\hypersetup{
colorlinks=true,
urlcolor= blue,
citecolor=blue,
linkcolor= blue,
}

\newif\ifptitle
\newif\ifpnumber
\newcounter{para}

\ptitletrue  
\pnumbertrue  

\newcommand{\mytitle}{Fitting the exotic hadron spectrum with an additional quark}

\begin{document}

\title{\mytitle}

\author{Scott Chapman}
\affiliation{Institute for Quantum Studies, Chapman University, Orange, CA  92866, USA}

\date{\today}

\begin{abstract}
Most of the exotic hidden-charm hadrons discovered over the last 20 years fit neatly into the quark model as normal mesons and baryons if the existence of a seventh flavor of quark is hypothesized. For the quark to reproduce the spectrum (mass, spin, parity) of exotic hadrons, it would have to have a mass of $\sim$2.9 GeV and a charge of $-\tfrac{1}{3}$.  The observed production and decay modes of these hadrons can be reproduced by a model that also includes light scalar bosons.      
\end{abstract}

\maketitle

\section*{Introduction}

In previous work, an alternative to the Standard Model was constructed from a form of broken supersymmetry that allows quarks and gauge bosons to be superpartners in a ``twisted superfield'' \cite{alternative,twisted}.  That relation allows the model to reproduce the symmetries and particles of the Standard Model with far fewer additional particles left over than in most supersymmetric models.  The Beyond-Standard-Model particles proposed by the model include an additional flavor of quark and some light scalar bosons. 

This paper shows that if there is an additional quark with charge $-\tfrac{1}{3}$ whose mass is approximately 2.9 GeV, then most of the observed exotic hadrons that decay to $c\bar{c}$ can be mapped to quark-model mesons and baryons involving the additional quark (rather than to 4- or 5-quark hadronic states).  The observed production and decay modes of these hadrons are reproduced by processes involving the scalars.

In the Standard Model, chiral anomalies generated by quark loops are cancelled by lepton loops, separately for each family.  With that anomaly cancellation method, it is not possible to add just one quark; an extension to the Standard Model must add two flavors of quark along with a family of leptons.  The model of \cite{alternative,twisted} cancels anomalies in a different way, separately for quarks and for leptons.  This is enabled by some quarks (and leptons) experiencing right-handed weak currents.  The experimental implications of this difference are discussed in \cite{twisted}, and it is shown that this structure not only reproduces well-established data, it also provides explanations for neutrino masses and for 3$\sigma$ anomalies in CKM $V_{us}$ data that hint at right-handed currents \cite{CKM-leptonic,Crivellin_2024}.

$R$ data ($\sigma(e^+e^-\to q\bar{q})/\sigma(e^+e^-\to \mu^+\mu^-)$) seemingly rule out an additional light quark.  If such a quark existed, it would generate an $R$ value too large to reproduce experimental data. In \cite{twisted}, it is shown that negative quantum corrections from the model's scalars reduce the $R$ value by roughly the same amount that the additional quark increases it, thereby allowing the model to reproduce $R$ data. That paper also points out structures in $R$ data that are difficult to explain in the Standard Model but are explained by this model.

\section{Hadron Mappings}

This section maps most of the known exotic hadrons as normal mesons and baryons involving the additional quark, denoted by the symbol $f$. 

According to the model of \cite{twisted}, production and decay processes of $f$-quark hadrons are primarily mediated by a light scalar $\varphi_5$ that does not interact with any gauge bosons, but does interact with certain quarks.  The scalar's main interactions are with $c\bar{c}$ and with $f(\bar{s}\cos\theta_C+\bar{d}\sin\theta_C )$, where $\theta_C$ is the Cabibbo angle.

Due to these interactions, $f$-hadrons can be generated any time a $c\bar{c}$ configuration is produced by an $e^+e^-$ collision or a $b$-hadron decay.  Scalar interactions can transform $c\bar{c}$ into either $f\bar{s}$ or $f\bar{d}$ (or their charge conjugates). 

$f$-hadrons can then decay via the reverse transformation or else the individual quark decay $f\to s+c\bar{c}$ or $f\to d+c\bar{c}$. Since the scalar is light, these decays are not suppressed like weak decays, helping to explain the width of $f$-hadrons.  In the next section, the observed production and decay of specific exotic hadrons is explained in detail.  In this section, the spectrum of observed exotic hadrons is mapped to the expected Quark-Model spectrum of $f$-quark mesons and baryons.

\subsection{f-quark mesons}

The following table maps observed exotic hadrons to $d\bar{f}$, $u\bar{f}$, and $s\bar{f}$ mesons.
\begin{equation}\label{fd1}
\begin{aligned}
&\textrm{Mapping of exotic hadrons to }f\textrm{-quark mesons}  \\
\renewcommand{\arraystretch}{1.5} 
&\begin{array}{|c|c|c|c|c|} \hline
\rm{QM} & d\bar{f} & u\bar{f} & s\bar{f} & \rm{Refs.} \\ 
\hline
1^1 S_0 & X^0(3250) & X^+(3250) &  & F|F|\,\,\, \\
1^3 S_1 & X^0(3350) &  & \psi(3760)  & F|\,\,\,|F \\
1^3 P_0 & X(3840) &  & X(3940) & \cite{3350}|\,\,\,|C\\
1^3 P_1 & \chi_{c1}(3872) & T_{cc}^+(3875) & \chi_{c1}(4010) & C|O|F\\1^1 P_1 & T_{c\bar{c}1}^0(3900) & T_{c\bar{c}1}^+(3900) & h_c(4000) & O|O|F \\
1^3 P_2 & \chi_{c0}(3915) &  &   & C|\,\,\,|\,\,\,\\
2^1 S_0 & Z_{cs}^0(3985) & Z_{cs}^+(3985) &  & \cite{39850}|\cite{3985}|\,\,\, \\
2^3 S_1 & T_{c\bar{c}}^0(4020)  & T_{c\bar{c}}^+(4020) & \psi(4230)  & O|O|C\\
1^1 D_2 &  & T_{c\bar{c}}^+(4055) & X(4160) &  \,\,\,|O|C\\
2^3 P_0 &  & T_{c\bar{c}}^+(4100) &  & \,\,\,|O|\,\,\,\\
2^3 P_1 & \chi_{c1}(4140)  &  & \chi_{c1}(4274)  & C|\,\,\,|C\\
2^1 P_1 &  & T_{c\bar{c}1}^+(4200) & h_c(4300) & \,\,\,|O|F \\
2^3 P_2 &  &  & X(4350) & \,\,\,|\,\,\,|C\\
1^3 D_1 & \psi(4320)   &  &   & \cite{4320}|\,\,\,|\,\,\,\\
3^3 S_1 & \psi(4390)  &  & \psi(4500) & \cite{4320}|\,\,\,|F \\
3^1 P_1 & & T_{c\bar{c}1}^+(4430) & T_{c\bar{c}\bar{s}1}^0(4600) & \,\,\,|O|\cite{LHCb_psi2Samp} \\
3^3 P_0 & \chi_{c0}(4500) & & \chi_{c0}(4700)  & C|\,\,\,|C\\
3^3 P_1 & &  & \chi_{c1}(4685)  & \,\,\,|\,\,\,|C\\
4^3S_1 & \psi(4660) & & Y(4710) & C |\,\,\,|\cite{4710} \\
\hline
\end{array}\,.
\end{aligned}
\end{equation}

In its 2025 update, Particle Data Group (PDG) lists integer-spin exotic hadrons in three places: (i) the ``$c\bar{c}$ Mesons'' list, (ii) the ``Other Mesons'' list, and (iii) a pdf describing ``Other Mesons: Further States'' \cite{PDG-mesons-2025}.  In the above table, the reference for each hadron is denoted as ``$C$'', ``$O$'', ``$F$'' for these lists, or else another reference if the hadron is not listed separately in PDG (e.g., it is grouped with another hadron).  All exotic hadrons in the ``$C$'' and ``$O$'' listings are represented in the above table, except for a few exceptions explained below. The meson masses in the above mapping consistently imply a mass of $2.8$ to $2.9$ GeV for the $f$ quark.

Some of the mappings stem from the following assumptions: $\psi(4360)$ is split into two resonances 4320 and 4390, as proposed in \cite{4320}.  $Y(4710)$ is not grouped together with the other measurements of $\psi(4660)$, as proposed in \cite{4710}.  $T_{c\bar{c}}^+(4050)$ and $T_{c\bar{c}}^+(4250)$ are not considered hadrons due to BaBar's non-observation. $T_{c\bar{c}0}^+(4240)$ (with $1^{+-}$ only disfavored by 1$\sigma$ relative to $0^{--}$) is assumed to be the same meson as $T_{c\bar{c}1}^+(4200)$. The LHCb observation of $X(3960)$ \cite{3960} is assumed to be the same as the PDG-listed $X(3940)$, and assumed to have $J^{PC}=0^{++}$.  The $0^{++}$ contribution near 3930 MeV seen in \cite{3915spin0} is assumed to come from the $X(3940)$ rather than the $\chi_{c0}(3915)$.  That weakens the $0^{++}$ case for $\chi_{c0}(3915)$ and allows its quantum numbers to be $2^{++}$, as assumed here.  The $X(3840)$ in the table is the 2.8$\sigma$ hint mentioned in \cite{3350}. $T_{cc}^+(3875)$ is discussed in the next section.

Many of the observed neutral mesons are assumed to be in the ``+'' $CP$ eigenstate.  For example, $\psi(4660)$ is assumed to be the $4^3 S_1$ meson of $(f\bar{d})_+=\tfrac{1}{\sqrt{2}}(f\bar{d}+d\bar{f})$, so it has $J^{PC}=1^{--}$.  The PDG-listed $X(4630)$ on the other hand, is mapped to the $4^3 S_1$ meson of $(f\bar{d})_-=\tfrac{1}{\sqrt{2}}(f\bar{d}-d\bar{f})$, which is the $1^{-+}$ $CP$ partner of $\psi(4660)$.

In PDG, the listing for $T_{c\bar{c}\bar{s}1}^+(4000)$ includes both a narrow resonance from BESIII and a wide resonance from LHCb.  Both the wide $T_{c\bar{c}\bar{s}1}^+(4000)$ and the $T_{c\bar{c}\bar{s}1}^+(4220)$ are assumed to be reflections from the following:  $B_c^+ \to X^+ \phi$, then $X^+ \to Z^+ J/\psi$, then $Z^+ \to K^+ K^0_L$, where $K^0_L$ is undetected.  For $m(X^+)\sim 5050$ MeV or $\sim 5300$ MeV (where $Z^+$ is $a_2(1320)$ or $\rho(1450)$), these decays look like the decays seen for the wide $T_{c\bar{c}\bar{s}1}^+(4000)$ or the $T_{c\bar{c}\bar{s}1}^+(4220)$.  In this model, the $X^+(5050)$ and $X^+(5300)$ are the $1^3S_1$ and $1^3P_1$ mesons of $c\bar{f}$.  The narrow $T_{c\bar{c}\bar{s}1}(4000)$ (aka $Z_{cs}(3985)$) is mapped in the table and is assumed to be the same as the $\eta_c(3945)$ of \cite{hc4000}.

The PDG-listed $T_{b\bar{s}}^+(5568)$ is the $2^3S_1$ meson of $c\bar{f}$.

The PDG-listed $T_{cc\bar{c}\bar{c}}^0(6900)$ are the $1^3P_J$ mesons of $f\bar{f}$.  Some may be prompt while others may come from the decay of $\Upsilon(nS)$.  Those from the latter will be mostly $1^3P_2$ with $J^{PC}=2^{++}$ for the same reason that $\Upsilon(nS)$ decays more often produce $\chi_{c2}$ than $\chi_{c1}$ or $\chi_{c0}$. The 6600 and 7100 peaks seen in \cite{CMSffspin} are the $1^3S_1$ and $2^1S_0$ mesons of $f\bar{f}$.  The 7100 may also have a contribution from a $2^3S_1$ $f\bar{f}$ meson with mass around 7250. The 6600 and 7250 decays are to $\gamma J/\psi J/\psi$, leaving $\sim 5/9$ of the remaining $J/\psi J/\psi$ configurations in the $2^{++}$ state.  The common $2^{++}$ configuration of these peaks matches with the configuration preferred by the CMS analysis \cite{CMSffspin}.  A mass of 7250 MeV for the $2^3S_1$ $f\bar{f}$ meson would be consistent with the 3.5$\sigma$ evidence of a resonance at that mass in $e^+e^-\to\mu^+\mu^-$ \cite{narrow_search}.  

\subsection{Other exotic mesons}

The model also has an octet of scalar fields $\varphi_1^a$ that have a mass similar to that of the charm quark when in an environment with energy density $\gtrsim$4 GeV/fm$^3$.  Below that scale by the Seiberg-Witten mechanism \cite{seiberg-witten}, the scalars become massless, form color monopoles, condense, and cause confinement, as described in \cite{twisted}.  Inside a $b$-hadron, the energy density is high enough that the hadron can decay to a meson along with a hadron formed of a colorless combination of two of these scalars.  In this model, the PDG-listed $T^{*0}_{cs0}(2870)$, $T^{*0}_{cs1}(2900)$ and $T^{*0}_{c\bar{s}0}(2900)$ are each di-scalars with either 0 or 1 unit of orbital angular momentum.  From \cite{twisted}, each of the scalars can decay to $c\bar{u}$ or $s\bar{d}$ or their conjugates, trading daughter quarks with the other scalar to form colorless mesons.  As a result, the di-scalar can decay to the observed daughter particles of $T^{*0}_{cs0}(2870)$, $T^{*0}_{cs1}(2900)$, and $T^{*0}_{c\bar{s}0}(2900)$.

In this model, the PDG-listed $T^{*++}_{c\bar{s}0}(2900)$ is assumed to be a reflection from a decay involving a doubly charged baryon that decays to mesons and a neutron, but the neutron is undetected.  One example of a scalar-mediated decay with the right kinematics is: $\Xi_{bc}^+\to \Xi_{cc}^{*++}(4000) D^-$ then $\Xi_{cc}^{*++}\to \Xi_{c}^{*+}(3055)\pi^+$ then $\Xi_{c}^{*+}\to D_s^+ n$, where the neutron is undetected. This relies on the $s\bar{d}\to d\bar{s}$ scalar octet interaction that is enabled inside hadrons with mass density $\gtrsim$4 GeV/fm$^3$.

In this model, the PDG-listed $T_{b\bar{b}1}(10610)$ and $T_{b\bar{b}1}^+(10650)$ are true 4-quark states, tetraquarks or meson molecules.

\subsection{f-quark baryons}

In this model, the observed positively charged pentaquarks are reinterpreted as isospin-1 $fuu$ baryons.  In the following table, the first 4 columns provide measured attributes of observed charged pentaquarks (masses and widths in MeV). The next two columns list some $suu$ baryons and their $J^P$.  The mapping assumes that the pentaquarks are actually $fuu$ baryons that are in the same quark-model state as the listed $suu$ baryons.  The last column shows the mass difference between the $fuu$ baryons and $suu$ baryons in the same state.
\begin{equation} \label{fuu} 
\begin{aligned}
&fuu\textrm{ baryons} \\
\renewcommand{\arraystretch}{1.5} 
&\begin{array}{|c|c|c|c|c|c|c|} \hline
\textrm{Name} & \rm{Mass} & \Gamma & J^P & suu\textrm{ Name} & J^P & \Delta m_s \\ 
\hline
P_c^+(4312) & 4312 & 10 & ?^? & \Sigma(1620) & {\tfrac{1}{2}}^- & 2692 \\
P_c^+(4380) & 4380 & 200 & ?^? & \Sigma(1660) & {\tfrac{1}{2}}^+ & 2720 \\
P^{N +}_{\psi}(4337) & 4337 & 29 & ?^? & \Sigma(1670) & {\tfrac{3}{2}}^- & 2675 \\
P_c^+(4440) & 4440 & 21 & ?^? & \Sigma(1750) & {\tfrac{1}{2}}^- & 2690 \\
P_c^+(4457) & 4457 & 6 & ?^? & \Sigma(1775) & {\tfrac{5}{2}}^- & 2682 \\
P_c^+(4457) & 4457 & 6 & ?^? & \Sigma(1780) & {\tfrac{3}{2}}^+ & 2677 \\
\hline
\end{array}\,
\end{aligned}
\end{equation} 
The data for $P^{N +}_{\psi}(4337)$ are from \cite{4337}, while those for the other pentaquarks are from \cite{PDG-mesons-2022}.  The last column shows mass differences between the observed pentaquarks and $suu$ baryons proposed to be in the same quark-model states.  The consistent mass differences imply that if the charged pentaquarks are actually $fuu$ baryons, the $f$ quark should have an effective mass of around 2.8 to 2.9 GeV.  

The reason that $P_c^+(4457)$ is listed twice in the table is because that pentaquark is mapped to a double peak $fuu$ analog of the $\Sigma(1775)$ and $\Sigma(1780)$.  

The next table maps the observed neutral pentaquarks to isospin 0 and isospin 1 baryon states of $fdu$.
\begin{equation} \label{fud} 
\begin{aligned}
&fdu\textrm{ baryons} \\
\renewcommand{\arraystretch}{1.5} 
&\begin{array}{|c|c|c|c|c|c|c|} \hline
\textrm{Name} & \rm{Mass} & \Gamma & J^P & sdu\textrm{ Name} & J^P & \Delta m_s \\ 
\hline
P^{\Lambda 0}_{\psi s}(4338) & 4338 & 7 & {\tfrac{1}{2}}^- & \Lambda(1670) & {\tfrac{1}{2}}^- & 2676 \\
P^{\Lambda 0}_{\psi s}(4459) & 4459 & 17 & ?^? & \Sigma(1775) & {\tfrac{5}{2}}^- & 2684 \\
P^{\Lambda 0}_{\psi s}(4459) & 4459 & 17 & ?^? & \Sigma(1780) & {\tfrac{3}{2}}^+ & 2679 \\
\hline
\end{array}\,,
\end{aligned}
\end{equation} 
It is proposed that $P_{cs}^0(4459)$ from \cite{Pcs} is the $fdu$ isospin partner of the $P_c^+(4457)$ resonance(s).  It is also proposed that $P^{\Lambda 0}_{\psi s}(4338)$ from \cite{4338} is an $fdu$ isospin singlet. Again, the mass differences are consistent with a 2.8 to 2.9 GeV quark.

\section{Production and Decay Processes}

The production and decay processes for $f$-hadrons are explained in the following subsections. Most of the production mechanisms involve either a virtual photon or a weak decay generating a $c\bar{c}$ that is then transformed to $f\bar{s}$ or $f\bar{d}$ via a scalar interaction.  Decays of $f$-hadrons to other $f$-hadrons are typically mediated by gluons or photons. Decays of $f$-hadrons to non-exotic hadrons are typically mediated by a light scalar or the $W$ boson.

\subsection{3872 and 3875} 

The $\chi_{c1}(3872)$ was first observed via the production $B^+\to K^+ \chi_{c1}$ and decay $\chi_{c1}\to \pi^+\pi^- J/\psi$ \cite{3872_2003}.  In this model, the $\chi_{c1}(3872)$ is an $(f\bar{d})_+=\tfrac{1}{\sqrt{2}}(f\bar{d}+d\bar{f})$ meson in the $1^3P_1$ $QM$ state.  The above production process involves the weak decay $u\bar{b}\to u\bar{c}c\bar{s}$ followed by the scalar-mediated process $c\bar{c}\to (f\bar{d})_+$. The majority of $\chi_{c1}\to \pi^+\pi^- J/\psi$ decays involve the reverse ``annihilation'' process $(f\bar{d})_+ \to c\bar{c}$ followed by emission of a $\rho^0$ or $\omega$ (either directly, or via emission of a virtual photon).

A smaller fraction of the observed decays involve the scalar-mediated quark decay $f\to s c\bar{c}$.  This mostly manifests as $\chi_{c1}(3872)\to K^0_S J/\psi$.  An excess at 500 MeV in the 2-pion invariant mass for these decays (fig. 9 of \cite{3872isospin}) suggests that this channel may account for up to 5-7\% of all $\chi_{c1}\to \pi^+\pi^- J/\psi$ decays.  

Like the annihilation decay, the scalar-mediated quark decay $f\to d c\bar{c}$ can also generate $\chi_{c1}\to \rho^0 J/\psi$.  This quark decay is Cabibbo suppressed relative to the $K^0_S$ decay, so its contribution should be smaller.  However, since it is an S-wave rather than P-wave decay, the quark-decay channel for $\chi_{c1}\to \rho^0 J/\psi$ could account for as much as 2-4\% of all $\chi_{c1}\to \pi^+\pi^- J/\psi$ decays.

In this model, the $\chi_{c1}(3872)$ has charged isospin partner mesons with quark content $u\bar{f}$ and $f\bar{u}$ that are also in the $1^3P_1$ $QM$ state.  In the table in the last section, the positively charged partner was mapped to $T_{cc}^+(3875)$, and justification is provided below for this mapping.  For now, until that mapping is motivated, the isospin partners of $\chi_{c1}(3872)$ will be denoted $T^\pm (3872)$.  The $u$ and $\bar{f}$ quarks in $T^+$ cannot annihilate to form $c\bar{c}$ like the first decay of $\chi_{c1}(3872)$ mentioned above.  The $T^+$ can, however, decay by $\bar{f}\to \bar{d} c\bar{c}$.  This generates $T^\pm (3872) \to \rho^\pm J/\psi$, but just as above, the branching ratio for this decay should only be 2-4\% of the branching ratio for $\chi_{c1}(3872)\to \pi^+\pi^- J/\psi$ decays.  This decay of a $T^\pm (3872)$ was searched for in \cite{3872isospin}.  No signal was found and a branching ratio upper limit of less than 50\% that of $\chi_{c1}(3872)\to \pi^+\pi^- J/\psi$ was established. Given the 2-4\% expectation above, that upper limit does not rule out this model.  In fact in \cite{3872isospin}, the largest event yield at a mass of 3873 had the right magnitude to potentially hint at the decay $T^\pm (3872) \to \rho^\pm J/\psi$.  

But there are other non-Cabibbo-suppressed decays $f\to s c\bar{c}$ in which it should be easier to observe $T^\pm (3872)$.  For example, the decay amplitude of $T^\pm\to K^\pm J/\psi$ should be similar to that of $\chi_{c1}(3872)\to K^0_S J/\psi$.  The former decay may be even easier to observe after a different production mode:  $B^+\to \phi T^+$.  This production starts with the weak decay $u\bar{b}\to u\bar{c}c\bar{s}$ followed by the scalar process $c\bar{c}\to \bar{f}s$ for a total process of $u\bar{b}\to u\bar{f}+s\bar{s}$.  Fig. 6c of \cite{Babarphi3872} may contain a hint that $T^+\to K^+ J/\psi$ is responsible for about 5\% of all $B^+\to \phi K^+ J/\psi$ decays.  

LHCb has studied this same decay with far more statistics than \cite{Babarphi3872} and has seen no evidence of the above decay \cite{Zcs4000}. As discussed in the previous section, it is proposed that some of the apparent LHCb $B^+\to \phi K^+ J/\psi$ decay data actually came from $B_c^+$ mesons decaying to $\phi K^+ J/\psi$ along with an undetected $K^0_L$.  It is proposed that the data from these $B_c^+$ decays obfuscate the signal from $T^\pm (3872)$.  This hypothesis could easily be tested by Belle II studying with more statistics $B^\pm\to \phi K^\pm J/\psi$ only for $B^\pm$ coming from $e^+e^-\to \Upsilon(4S)\to B^+B^-$. These events cannot involve $B_c^+$ mesons.

The decay $T^\pm(3872)\to K^\pm J/\psi$ could potentially also be seen after the scalar-mediated production mode $\psi(4660)\to K^{\mp} T^\pm$.  Fig. 7a in the supplemental material of \cite{4710} may contain a hint of this decay as well as the related decay $T^\pm\to K^{*\pm}(700) J/\psi$ with a $\pi^0$ unobserved.

Another way to see $T^+ (3872)$ again involves the scalar-mediated decay $u\bar{f}\to u\bar{s} c\bar{c}$, but this time the quarks decay to $p\bar{\Lambda}$, similarly to the way that $\chi_{c1}(1P)$ decays to $p\bar{p}$.  Incidentally, $X^+(3840)$ (the proposed isospin partner of $X(3840)$ in the first table) could also decay in this way.  Fig. 1 of \cite{Belle20093875p} may contain a hint of these predicted decays; the paper mentions an unexplained excess there. 

In the section above describing this model's explanation of the $T^{*0}_{cs0}(2870)$, $T^{*0}_{cs1}(2900)$ and $T^{*0}_{c\bar{s}0}(2900)$ hadrons, it was mentioned that this model has an octet of scalars $\varphi_1^a$ that condense below an energy density of about 4 GeV/fm$^3$.  But for hadrons with masses over about 3.5-3.8 GeV, these scalars can participate in decay processes. Their interactions are fully described in \cite{twisted}, but one aspect is particularly interesting for the $T_{cc}^+(3875)$.  Unlike gluon-mediated interactions, $\varphi_1^a$-mediated interactions do not conserve charm.

Two of the interaction terms for $\varphi_1^a$ are:
\begin{equation}\label{octetg}
\varphi_1^{\dag a}\left(\tfrac{1}{\sqrt{2}}g_3 \bar{c}'_L t^a u'_R +\Gamma_\Phi\bar{f}'_Rt^a s'_L +...\right)+h.c.,
\end{equation} 
where $g_3$ is the strong coupling, $t^a$ are half the Gell-Mann matrices, $\Gamma_\Phi$ is a superpotential coupling, and primes on the quark fields denote that they are gauge rather than mass eigenstates.  To the latter point, roughly 22\% of $s'_L$ is $d_L$ (Cabibbo mixing).  Interactions with $\varphi_1^{\dag a}$ can therefore create $c\bar{u}$ and change $\bar{f}\to\bar{d}$.  There are terms in the superpotential with three factors of $\varphi_1^{\dag a}$, and those terms can enable the following decay: $\bar{f}\to\bar{d}c\bar{u}c\bar{u}$. 

In this model, the $T^+(3872)$ is a $u\bar{f}$ meson.  The above mechanism therefore enables the decays $T^+(3872)\to \pi^+D^0D^0$ or $T^+(3872)\to \pi^0D^+D^0$.  In other words, it can reproduce the decays observed for $T_{cc}^+(3875)$.  This is the motivation for the proposal in this model that $T_{cc}^+(3875)$ is the $T^+(3872)$, the isospin partner of $\chi_{c1}(3872)$.  A consequence of this proposal is that $\chi_{c1}(3872)$ should also have decays to $\pi^0D^0D^0$ and $\pi^0\bar{D}^0\bar{D}^0$.

\subsection{Vector mesons}

One way to produce the $J^{PC}=1^{--}$ vector meson $(f\bar{s})_+$ and $(f\bar{d})_+$ $CP$ eigenstates is via $e^+e^-$ collisions.  The process is $e^+e^- \to c\bar{c} \to (f\bar{s})_+$ or $(f\bar{d})_+$, where a photon mediates the first part and a scalar mediates the second part.  The $(f\bar{d})_+$ process is Cabibbo suppressed relative to the $(f\bar{s})_+$ process, so the latter should be much larger resonances.

With that in mind, the three major resonances seen in \cite{4710} motivate the mapping of $\psi(4230)$, $Y(4500)$ and $Y(4710)$ to the $2^3S_1$, $3^3S_1$ and $4^3S_1$ resonances of $(f\bar{s})_+$. Their decay to $K^+K^- J/\psi$ is facilitated by either $f\bar{s}\to s\bar{s}c\bar{c}$ or the annihilation process $(f\bar{s})_+\to c\bar{c}$, neither of which are Cabibbo suppressed. 

The quark decay $(f\bar{s})_+\to s\bar{s} + c\bar{c}$ can generate $f_0 J/\psi$, with the $f_0$ decaying to $\pi^+\pi^-$ or $K^+K^-$. $(f\bar{s})_+$ mesons can also generate $\pi^+\pi^- J/\psi$ after the $(f\bar{s})_+\to c\bar{c}$ annihilation process.  These are both assumed to occur for $R(3760)$ \cite{3760}, the $1^3S_1$ $(f\bar{s})_+$ vector meson.

The $(f\bar{d})_+$ vector mesons should generate smaller resonances in $e^+e^-$ collisions.  Recent amplitude analyses of BESIII data suggest the possibility that the $\psi(4360)$ is actually two separate resonances: $\psi(4320)$ and $\psi(4390)$ with masses around 4294 and 4406 (Model I) \cite{4320}.  Making that assumption, these resonances are mapped to the $1^3D_1$ and $3^3S_1$ resonances of $(f\bar{d})_+$.  That split mapping helps with the following subtlety.

In this model, there is no gluon-mediated decay from the $(f\bar{s})_+$ meson $\psi(4230)$ to $\pi^\mp T_{c\bar{c}1}^\pm(3900)$.  However, there is a gluon-mediated decay from the $(f\bar{d})_+$ meson $\psi(4320)\to \pi^\mp T_{c\bar{c}1}^\pm(3900)$, since $T_{c\bar{c}1}^+(3900)$ is a $u\bar{f}$ meson.  With a mass 80 MeV smaller than the one usually assigned to $\psi(4360)$ and a large width, the $\psi(4320)$ is better able to produce the observed decays to $T_{c\bar{c}1}^\pm(3900)$.  

By a similar argument, it is assumed that the radiative decay to $\chi_{c1}(3872)$ that has been attributed to $\psi(4230)$ is actually the decay $\psi(4320)\to\gamma\chi_{c1}(3872)$.

The $T_{c\bar{c}}^{\pm,0}(4020)$ are mapped in this model to the $2^3S_1$ mesons of $f\bar{u}$, $u\bar{f}$, $f\bar{d}$ and $d\bar{f}$.  The neutral meson $(f\bar{d})_+$ could be responsible for the enhanced cross section near $\sqrt{s}=4$ GeV in \cite{4320}.

The final vector meson mapping is of the observed $X^0(3350)$ \cite{3350} to the $1^3S_1$ meson of $(f\bar{d})_+$.  Gluons or scalars mediate the decay processes of all of the other mesons in this section.  But an $(f\bar{d})_+$ vector meson can also decay by the $W$-mediated process $f\bar{d}\to c\bar{d}d\bar{u}$.  This is the process assumed for the observed decay $X^0(3350)\to \Lambda_c \bar{p}$.

\subsection{Other mesons}

In this model, the $1^1S_0$ meson of $(f\bar{d})_+$ is the $X^0(3250)$ with $J^{PC}=0^{-+}$. In a scalar-mediated annihilation decay, the resulting $c\bar{c}$ configuration has the same quantum numbers as $\eta_c$, which is known to decay to $K^+\bar{p}\Lambda$.  Therefore the decay $X^0(3250)\to K^+\bar{p}\Lambda$ is enabled by this model, and those decays have been observed \cite{3250,PDG-mesons-2022}.  The $X^\pm(3250)$ $u\bar{f}$ and $f\bar{u}$ mesons that decay by emitting a $\pi^\pm$ before the above $c\bar{c}$ decays have also been observed.

As mentioned in the mapping section, the $X(3960)$ is mapped to the $1^3P_0$ $0^{++}$ meson of $(f\bar{s})_+$.  It has been observed decaying to $D_s^+D_s^-$.  In this model, that is mediated by the scalar process $(f\bar{s})_+\to s\bar{s}c\bar{c}$.

An $f\bar{f}$ meson in this model can decay to $s\bar{f}c\bar{c}$ then $c\bar{c}c\bar{c}$, with scalars mediating both processes.  This is what is assumed to occur in the observed $T_{cc\bar{c}\bar{c}}(6900)\to J/\psi J/\psi$ decay, where $T_{cc\bar{c}\bar{c}}(6900)$ is mapped to the $1^3P_J$ mesons of $f\bar{f}$. 

According to the model, the $1^3S_1$ meson of $f\bar{f}$ should have a mass around 6600 MeV.  It would then have enough mass to decay to two $X(3250)$ mesons.  Since that decay is not OZI suppressed, the Y(6600) should be a wide resonance detectable in $e^+e^-\to$ hadrons.  The Mark 1 data shown in fig 10 of \cite{lowR2} may contain evidence of that resonance.

\subsection{Baryons}

The $f$-quark baryons of this model are mapped to the observed pentaquarks.  One observed process for pentaquark production is $\Lambda_b\to P_c^+K^-$.  In this model, $P_c^+$ is an $fuu$ baryon.  Production is initiated by the $W$-mediated decay $bud\to cud\bar{c}s$.  This is followed by a scalar-mediated process $c\bar{c}\to f\bar{d}$ and a gluon-mediated process $d\bar{d}\to u\bar{u}$.  The net result is the process $bud\to fuu+s\bar{u}$.  The decay $P_c^+\to p J/\psi$ is enabled by the scalar-mediated decay $f\to dc\bar{c}$. 

There is evidence for the process $\bar{B}_s^0\to P_\psi^{N+}\bar{p}$.  In this model, $P_\psi^{N+}$ is also an $fuu$ baryon.  Production is initiated by the $W$-mediated decay $b\bar{s}\to c\bar{s}\bar{c}s$.  This is followed by a scalar-mediated process $c\bar{c}\to f\bar{d}$ and a gluon-mediated process $s\bar{s}\to u\bar{u}u\bar{u}$.  The net result is the process $b\bar{s}\to fuu+\bar{u}\bar{u}\bar{d}$.  The decay $P_\psi^{N+}\to p J/\psi$ is enabled by the scalar-mediated decay $f\to dc\bar{c}$.

Another observed process is $B^-\to P_{\psi s}^{\Lambda 0}\bar{p}$. In this model, $P_{\psi s}^{\Lambda 0}$ is an isospin-0 baryon of $fdu$. Production is initiated by the $W$-mediated decay $b\bar{u}\to c\bar{u}\bar{c}s$.  This is followed by a scalar-mediated process $c\bar{c}\to f\bar{s}$ and a gluon-mediated process $s\bar{s}\to u\bar{u}d\bar{d}$.  The net result is the process $b\bar{u}\to fdu+\bar{u}\bar{u}\bar{d}$.  The decay $P_{\psi s}^{\Lambda 0}\to \Lambda J/\psi$ is enabled by the scalar-mediated decay $f\to sc\bar{c}$.

\section{Unconfirmed Hints of Heavier Exotic Hadrons}

It is possible that evidence for an $f\bar{c}$ meson has been seen. Namely, evidence for a narrow resonance in $\pi^-B_s^0$ at 5568 MeV was reported in \cite{5568} but was not reproduced in other experiments that had different CM energies or kinematics.  Given the mass of the $f$ quark in this model, the $T_{b\bar{s}}^-(5568)$ has a mass that would make it an attractive candidate for the $2^3S_1$ state of $f\bar{c}$. Other potential hints of $c\bar{f}$ mesons were described in the mapping section above.

It is also possible that evidence for an $(f\bar{b})_+$ vector meson was seen in a past experiment.   Given the mass of the $f$ quark in this model, the $5^3S_1$ meson of $(f\bar{b})_+$ could have a mass just slightly larger than that of $\Upsilon(1S)$.  This $1^{--}$ $(f\bar{b})_+$ meson could decay by gamma emission to the $1^3P_1$ meson of $(f\bar{b})_+$.  The latter meson could have a mass near 8.3 GeV.

The Crystal Ball collaboration reported evidence from their 1983 data that the $\Upsilon(1S)$ had a radiative decay to a narrow resonance at 8322 MeV \cite{CB1983}.  The resonance was not seen in 1984 Crystal Ball data or in those of other experiments with $e^+e^-$ centered on the $\Upsilon(1S)$ resonance \cite{CB1984}.  However it was noticed that $R_{\rm hadron}$ for the original dataset was smaller than $R_{\rm hadron}$ for the other $\Upsilon(1S)$ datasets.  This suggested that it was possible that the mean CM energy of the 1983 Crystal Ball data may have been a bit higher than the $\Upsilon(1S)$ resonance.  If so, and if there was another $1^{--}$ resonance just a little bit more massive than the $\Upsilon(1S)$, then the 1983 resonance could have been real, not an experimental fluke.  It was stated that ``the 1984 result rules out the 1983 result to the 90\% C.L. level unless some state exists between about 16-26 MeV above the Upsilon(1S)'' \cite{CBR_Had}.  That possibility was thought to be very unlikely, so it was assumed that the 1983 evidence of a resonance and smaller $R_{\rm hadron}$ value were both experimental flukes that should be ignored.

This model opens the door to the possibility that the 8322 MeV resonance in 1983 data was not a fluke.  As mentioned above, the $5^3S_1$ meson of $(f\bar{b})_+$ could have a mass a little larger than that of the $\Upsilon(1S)$, and it could decay radiatively to the $1^3P_1$ meson of $(f\bar{b})_+$, producing the resonance seen in the 1983 data.  The other experiments would not have seen that resonance since their CM energies were centered on $\Upsilon(1S)$, rather than a little above it. 

Other potential hints of $f\bar{b}$ mesons are described in \cite{exotics}.

\section*{Conclusion}

Over the last twenty years, many dozens of exotic hadrons have been observed.  It is generally thought that these are 4- or 5-quark combinations of the known Standard-Model quarks.  Some theories propose structures like tetraquarks and pentaquarks.  Others propose meson molecules, hadrocharmonium, hybrids or more exotic structures.  None of the theories can reproduce all of the data, and they all struggle with explaining the burgeoning spectrum of observed exotic hadrons.

This work has shown that the spectrum of masses, spins, parity and other quantum numbers of the exotic hadrons are fit extremely well using just mesons and baryons involving an additional flavor of quark.  The theoretical background for this additional quark is developed in \cite{twisted}.  The light scalars in that theory provide a simple explanation for all of the observed production and decay processes.

If this model is correct, then it should be possible to predict the existence of many more exotic hadrons as well as new decays of already-observed exotic hadrons.  Dozens of such predictions are made in \cite{exotics,HADRON2025,LHCbpres2025} for new hadrons or new decays that could be observed in current experiments.

\end{document}